# Glassy atomic vibrations and blurry electronic structures created by local structural disorders in high-entropy metal telluride superconductors

Yoshikazu Mizuguchi[1]*, Hidetomo Usui[2], Rei Kurita[1], Kyohei Takae[3], Md. Riad Kasem[1], Ryo Matsumoto[4], Kazuki Yamane[4,5], Yoshihiko Takano[4,5], Yuki Nakahira[1,6], Aichi Yamashita[1], Yosuke Goto[1,7], Akira Miura[8], Chikako Moriyoshi[9]

1. Department of Physics, Tokyo Metropolitan University, Hachioji 192-0397, Japan.
2. Department of Physics and Materials Science, Shimane University, Matsue, Shimane 690-8504, Japan
3. Department of Fundamental Engineering, Institute of Industrial Science, University of Tokyo, Meguro-ku, Tokyo 153-8505, Japan
4. International Center for Materials Nanoarchitectonics (MANA), National Institute for Materials Science, Tsukuba, Ibaraki 305-0047, Japan
5. University of Tsukuba, Tsukuba, Ibaraki 305-8577, Japan
6. Quantum Beam Science Research Directorate, National Institutes for Quantum Science and Technology, Hyogo 679-5148, Japan
7. National Institute of Advanced Industrial Science and Technology (AIST), Tsukuba, Ibaraki 305-8568, Japan
8. Faculty of Engineering, Hokkaido University, Sapporo 060-8628, Japan
9. Graduate School of Advanced Science and Engineering, Hiroshima University, Higashihiroshima, Hiroshima, 739-8526, Japan

* Corresponding author: Y. Mizuguchi (mizugu@tmu.ac.jp)

**Abstract**

Recently, high-entropy (HE) metal chalcogenides have been extensively studied as high-performance thermoelectric materials or exotic superconducting materials. The motivation of this work is our recent observation of the robustness of superconductivity in a HE superconductor $Ag_{0.2}In_{0.2}Sn_{0.2}Pb_{0.2}Bi_{0.2}Te$ (CsCl-type) to external pressure. The superconducting transition temperature ($T_c$) of $Ag_{0.2}In_{0.2}Sn_{0.2}Pb_{0.2}Bi_{0.2}Te$ is almost constant with pressure, described as robustness of superconductivity to pressure, whereas the PbTe with zero configurational entropy of mixing exhibits a clear decrease in $T_c$ with pressure. Here, we investigated the atomic displacement parameters ($U_{iso}$), the atomic-vibration characteristics, and the electronic states of metal tellurides (MTe) with various configurational entropy of mixing ($\Delta S_{mix}$) at the M site. The $U_{iso}$ for the M site is clearly increased by M-site alloying with $\Delta S_{mix} \geq 1.1R$, which is the evidence of local disorder introduced by the increase in $\Delta S_{mix}$ via the solution of three or more M elements. The revealed vibrational density of states (DOS) shows a remarkable broadening with $\Delta S_{mix} \geq 1.1R$, which indicates glassy characteristics of atomic vibration in HE MTe with a NaCl-type structure (low-pressure phase). On the electronic states of the CsCl-type (high-pressure) phases, where the robustness of $T_c$ is observed, blurry electronic band structure appears with increasing $\Delta S_{mix}$, which indicates the evolution of blurry (glassy) electronic states in HE MTe with the CsCl-type structure. The estimated electronic DOS at Fermi energy cannot explain the changes in $T_c$ for HE MTe when assuming conventional electron-phonon superconductivity, but the conventional explanation seems to work for PbTe. Therefore, the pairing mechanisms in MTe with $\Delta S_{mix} \geq 1.1R$ are affected by glassy phonon and/or blurry electronic states in MTe, and the robustness of superconductivity would be originating from unique electron-phonon coupling.

# 1. Introduction

Disorder in functional materials is one of the parameters essential for the enhancement of properties of functional materials. For example, that can suppress thermal conductivity in thermoelectric materials, which is important to improve dimensionless figure-of-merit ($ZT$) [1–4]. In addition, disordered superconducting materials have been studied in the field of pure and applied physics of superconductivity [5–10]. Despite the importance of disorders in functional materials, evaluation of structural disorder and clarification of the effects of the disorders on electronic and phonon characteristics had been generally challenging because of the difficulty of inclusion of the effects of disorder in theoretical calculations.

Recently, new class of disordered functional materials have been designed with the concept of high-entropy alloy (HEA), which is an alloy containing five or more elements solving in a crystallographic site with a concentration of 5–35 at% [11]. The criterion satisfies a high configurational entropy of mixing ($\Delta S_{mix}$) higher than $1.5R$, where $R$ is gas constant; $\Delta S_{mix}$ is defined as $\Delta S_{mix} = -R \Sigma_i c_i \ln c_i$, where $c_i$ is composition of the component $i$. In the fields of alloy engineering, simple alloy-based materials have been extensively studied [11,12]. In addition, superconductivity has been observed in various HEAs [13–15]. It is worth noting that the superconducting properties of HEA superconductors are different from those of pure-crystalline and amorphous superconductors [14]. In addition, there are reports on the relationship between the SC properties of HEAs and mechanical properties [16,17]. These facts suggest that the local structures and/or disorders play important roles to determine their physical properties.

Since 2018, HEA-type *compound* superconductors, for example, layered systems ($REO_{0.5}F_{0.5}BiS_2$ and $REBa_2Cu_3O_{7-d}$; RE: solution of rare-earth elements) [18–22] and NaCl-type metal chalcogenides MCh (M = Ag, In, Ge, Sn, Sb, Bi, Pb; Ch = S, Se, Te) [23–36], have been developed. In those HEA-type compounds, one or more crystallographic sites are alloyed, which results in $\Delta S_{mix} > 1.5R$ and satisfies the criterion of HEA [27]. Among them, a notable SC property was observed in a HEA-type MTe superconductor, $Ag_{0.2}In_{0.2}Sn_{0.2}Pb_{0.2}Bi_{0.2}Te$ (AgInSnPbBiTe$_5$); robustness of superconductivity under high pressure was observed [28], which is similar to that observed in an alloy-based HEA Ti-Zr-Hf-Nb-Ta [29]. The robustness is discussed as how the $T_c$ is insensitive to the changes of pressure (increase in pressure); hence, the pressure differential of $T_c$ ($dT_c/dP$) is the scale of the phenomena. For Ti-Zr-Hf-Nb-Ta, the origin of the robustness of superconductivity under high pressure was proposed as the robustness of electronic density of states under high pressure [30]. In HEA-type MTe, however, the pressure dependences of electronic structure examined by X-ray absorption spectroscopy for HEA-type $Ag_{0.2}In_{0.2}Sn_{0.2}Pb_{0.2}Bi_{0.2}Te$ were quite similar to those for PbTe, which does not exhibit robustness of superconductivity under high pressure and shows a decrease in $T_c$ [28]. Therefore, we consider that the mechanisms of the robust superconductivity under high pressures in HEA-type MTe are not simply explained by electronic structure only and should be related to local structural disorder introduced by M-site alloying. Hence, we decided to investigate the effects of M-site alloying on atomic displacements (static local disorders), atomic vibrations, electronic structure, and superconductivity. We observed glassy vibrational density of states (VDOS) and blurry electronic band structure in MTe with $\Delta S_{mix} \geq 1.1R$ by the molecular dynamics (MD) simulations and the band calculations. Here, we described the *glassy* atomic vibration as the characteristics with flat VDOS with high vibrational entropy but with no atomic dispersions. On the basis of the results, we propose that the robustness of superconductivity under high pressures is caused by glassy atomic vibrations, blurry electronic structures, and unique electron-phonon coupling in HE MTe. The results shown here would be useful for developing new HE functional (electronic) materials including superconductors and thermoelectric materials [2–4].

## 2. Materials and methods

The polycrystalline samples of PbTe ($\Delta S_{mix} = 0$), $Sn_{0.5}Pb_{0.5}Te$ ($\Delta S_{mix} = 0.69R$), $Ag_{1/3}Pb_{1/3}Bi_{1/3}Te$ ($\Delta S_{mix} = 1.10R$), $Ag_{0.2}In_{0.2}Sn_{0.2}Pb_{0.2}Bi_{0.2}Te$ ($\Delta S_{mix} = 1.61R$, HEA-1), and $Ag_{0.175}In_{0.3}Sn_{0.175}Pb_{0.175}Bi_{0.175}Te$ ($\Delta S_{mix} = 1.58R$, HEA-2) were prepared by solid-state reaction in an evacuated quartz tube. For the HEA-type (Ag,In,Sn,Pb,Bi)Te samples, high-pressure annealing was performed to obtain the NaCl-type single phase. Details of sample preparation have been reported in Refs. 23, 26, and 28.

Temperature-dependent synchrotron X-ray diffraction (SXRD) experiments were performed at the beamline BL02B2, SPring-8. Typical SXRD patterns and the estimated lattice constants are shown in Fig. S1. The wavelength in the experiments for PbTe, $Sn_{0.5}Pb_{0.5}Te$, and $Ag_{1/3}Pb_{1/3}Bi_{1/3}Te$ was 0.495559(1) Å (proposal No. 2020A1096), and that for $Ag_{0.2}In_{0.2}Sn_{0.2}Pb_{0.2}Bi_{0.2}Te$ was 0.495259(1) Å (proposal No. 2020A0068). The SXRD experiments were performed with a sample rotator system; the diffraction data were collected using a high-resolution one-dimensional semiconductor detector (multiple MYTHEN system [31]) with a step size of $2\theta = 0.006°$. The samples were packed in an evacuated quartz capillary with an inner diameter of 0.1 mm, and the temperature of the sample was controlled by $N_2$ gas. The obtained SXRD patterns were refined by the Rietveld method using RIETAN-FP [32]. The schematic images of the crystal structures were drawn by VESTA [33].

The temperature ($T$) dependences of electrical resistance ($R$) under pressures ($P$) were measured on a Physical Property Measurement System (PPMS, Quantum Design) using an originally designed diamond anvil cell (DAC) with boron-doped diamond electrodes [34–36]. The sample was placed on the boron-doped diamond electrodes in the center of the bottom anvil. The surface of the bottom anvil, except for the sample space and electrical terminal, was covered with the undoped diamond insulating layer. The cubic boron nitride powders with ruby manometer were used as a pressure-transmitting medium. The applied pressure was estimated by the fluorescence from ruby powders [37] and the Raman spectrum from the culet of top diamond anvil [38] using an inVia Raman microscope (RENISHAW). The definition of $T_c$ from the onset of superconducting transition is described in Fig. S2(c).

In this study, we performed MD simulation to investigate the vibration of the particles (atoms). Particles $i$ and $j$ interact via a Lennard-Jones (LJ) potential and a Coulomb potential.

$$m\ddot{\boldsymbol{r}}_i = \sum_{j \neq i} \boldsymbol{F}(\boldsymbol{r}_{ij}),$$

$$\boldsymbol{F}(\boldsymbol{r}_{ij}) = -\partial U_{\text{LJ}}/\partial \boldsymbol{r}_{ij} + \boldsymbol{F}_q,$$

$$U_{\text{LJ}} = 4\varepsilon \left[ \left(\frac{\sigma_{ij}}{r_{ij}}\right)^{12} - \left(\frac{\sigma_{ij}}{r_{ij}}\right)^6 \right],$$

$$\boldsymbol{F}_q = k q_i q_j \boldsymbol{r}_{ij}/r_{ij}^3,$$

where $\boldsymbol{r}_{ij}$ is center-to-center displacement vector from particle $j$ to $i$ and $r_{ij}$ is its absolute value, $m$ is the mass common to all the particles, $\varepsilon$ is the coefficient for LJ potential, and $\sigma_{ij} = (\sigma_i + \sigma_j)/2$ with $\sigma_i$ being the size of the particle $i$. $k$ is the Coulomb constant and $q_i$ is the charge of the particle $i$. The length of the particles is normalized by the ionic diameter of Te$^{2-}$ ($\sigma_{\text{Te}}$=2.21 Å). Then, the particle size of Ag, In, Sn, Pb, and Bi become 0.520, 0.362, 0.421, 0.538, and 0.466, respectively. Here, the atomic mass $m$ = 127.6 g/mol and the interatomic interaction $\varepsilon$ = 292 $k_B$ J are assumed to be constant to elucidate the role of size and charge dispersity. Thus, the time, temperature, and pressure are measured in units of $t_0 = \sqrt{m\sigma_{\text{Te}}/\varepsilon}$, $T_0 = \varepsilon/k_B$, and $P_0 = \varepsilon/\sigma_{\text{Te}}^3$, respectively. We arranged 6912 cation particles randomly and 6912 Te$^{2-}$ particles in a cubic structure. That is, the number of the unit cell is 12 in one direction. The cutoff length of LJ interaction is 1.13$\sigma_{\text{Te}}$. We computed $F_q$ by using the Ewald summation method. By choosing the Ewald parameter $\alpha$ = 0.6, the cutoff length in the real space $5\sigma_{\text{Te}}$ and the cutoff wavenumber $7(2\pi/L)$ where $L$ is the system length, the root mean square error in the force becomes less than $10^{-4}$. We prepared the initial condition by running 10000 MD steps under NPT ensemble, where the pressure tensor $P$ and temperature $T$ are controlled by Parrinello--Rahman's method and Nose--Hoover's thermostat, respectively [39]. The external pressure tensor is $P = P_0 \mathbf{1}$ in this study, where $\mathbf{1}$ is the unit tensor. Then we switched to NVE ensemble to compute VDOS $D(\omega)$. We calculate the VDOS by the Fourier transformation of the velocity autocorrelation function as

$$D(\omega) = \int_{-\infty}^{\infty} \frac{dt}{\pi} \frac{\langle \sum_j m\, v_j(t) \cdot v_j(0) \rangle}{\langle \sum_j m v_j(0)^2 \rangle} \exp(i\omega t),$$

where the bracket represents statistical and time average, and $D(\omega)$ is normalized to satisfy $\int_0^\infty d\omega D(\omega) = 1$. Then, the difference in the vibrational entropy per particle is computed by

$\Delta S/k_B = -\frac{N_f}{N} \int_0^\infty d\omega [D(\omega) - D_0(\omega)] \ln \omega$, where $N = 13824$ is the total number of the atoms, $N_f = 3N - 3$ is the degree of freedom, and $D_0(\omega)$ is the VDOS of PbTe at $T = 0.2$.

The electronic band structure of CsCl-type MTe was calculated using the Korringa-Kohn-Rostoker Green's function method with inclusion of spin-orbit coupling, as implemented in the AkaiKKR package [40]. Moreover, in order to treat disordered phase, the coherent potential approximation was used [41]. The generalized gradient approximation parameterized by Perdew-Burke-Ernzerhof [42] was adapted, and spin orbit coupling was considered for the calculations. A $k$-mesh was set to 13×13×13, and the width of the energy contour for complex integration was set to 2.1 Ry for (Ag,In,Sn,Pb,Bi)Te, and 1.5 Ry for PbTe.

3. Results

Since clear differences in the $T_c$-$P$ trend have been observed in MTe (M = Pb, $Ag_{1/3}Pb_{1/3}Bi_{1/3}$, $Ag_{0.2}In_{0.2}Sn_{0.2}Pb_{0.2}Bi_{0.2}$) in Ref. 28, we started this study with the experiments to further confirm the trend that the robustness of superconductivity to external pressure is induced by the increase in $\Delta S_{mix}$. For that, $R$-$T$ for $Sn_{0.5}Pb_{0.5}Te$ ($\Delta S_{mix} = 0.69R$) and In-rich $Ag_{0.175}In_{0.3}Sn_{0.175}Pb_{0.175}Bi_{0.175}Te$ ($\Delta S_{mix} = 1.58R$) were measured under high pressures. As reported in Ref. 26, the ambient-pressure phase of $Ag_{0.175}In_{0.3}Sn_{0.175}Pb_{0.175}Bi_{0.175}Te$ has the highest $T_c$ among the (Ag,In,Sn,Pb,Bi)Te phases because of the optimized In concentration.

The $T_c$-$P$ data are plotted in Fig. 2(a) with the data published in Ref. 28. The $R$-$T$ data are summarized in supporting information. Figure S2(a) shows the $R$-$T$ for $Sn_{0.5}Pb_{0.5}Te$ under high pressures. The $R$ for the low-$P$ phase is low, but it increases at $P = 5.4$ GPa and gradually decreases with pressure. The onset of superconductivity was observed at $P = 14.9$ GPa (see Fig. S2(c) for estimation of onset temperature), and the highest $T_c$ of 6.39 K was observed at $P = 21.2$ GPa. Then, the $T_c$ gradually decreases with pressure for $Sn_{0.5}Pb_{0.5}Te$. Figure S3 shows the $R$-$T$ for $Ag_{0.175}In_{0.3}Sn_{0.175}Pb_{0.175}Bi_{0.175}Te$ under high pressures. In the middle pressure region, superconductivity was suppressed, which is similar to $Ag_{0.2}In_{0.2}Sn_{0.2}Pb_{0.2}Bi_{0.2}Te$ [28], and a superconducting transition was reobserved at $P = 12.5$ GPa. Above 20 GPa, the $P$ dependence of $T_c$ exhibits robustness under high pressures. Noticeably, the $T_c$s for $Ag_{0.2}In_{0.2}Sn_{0.2}Pb_{0.2}Bi_{0.2}Te$ (HEA-1 in Fig. 2(a)) and $Ag_{0.175}In_{0.3}Sn_{0.175}Pb_{0.175}Bi_{0.175}Te$ (HEA-2 in Fig. 2(a)) under high pressures are almost the same, while $T_c$ at ambient pressure is clearly different [26].

Figure 2(b) shows the $dT_c/dP$ plotted as a function of $\Delta S_{mix}/R$. The $dT_c/dP$ was estimated by linear fitting as displayed in Fig. S4. By taking $dT_c/dP$, the trend that the robust superconductivity is induced when $\Delta S_{mix}$ increased has been evidently confirmed. Basically, in MTe, electronic DOS in the CsCl-type structure decreases with pressure (Fig. 5(d)). Therefore, the results observed for PbTe and $Sn_{0.5}Pb_{0.5}Te$ are well explained from the changes in electronic DOS. However, for $Ag_{1/3}Pb_{1/3}Bi_{1/3}Te$ ($\Delta S_{mix} = 1.10R$), the trend obviously changes, and the robustness of superconductivity to pressure appears: the $dT_c/dP$ approaches zero. For HEA-1 and HEA-2, the $dT_c/dP$ is close to zero. The results suggest that there is a clear difference in superconducting states between PbTe (or $Sn_{0.5}Pb_{0.5}Te$) and $Ag_{1/3}Pb_{1/3}Bi_{1/3}Te$. This fact motivated us to computationally investigate vibrational and electronic characteristics of NaCl-type (low-$P$) and CsCl-type (high-$P$) phases of MTe. Let us mention that single crystals of HE MTe phases are not available so far, which is the reason why we performed X-ray diffraction and computational investigations here.

We used synchrotron XRD to examine the local structural disorder (inhomogeneity of local crystal structure) in MTe with various M-site $\Delta S_{mix}$. In Fig. 3, the temperature dependences of isotropic displacement parameters ($U_{iso}$) for the M and Te sites are summarized. For PbTe and $Sn_{0.5}Pb_{0.5}Te$ ($\Delta S_{mix} = 0.69R$), as shown in Figs. 3(a) and 3(b), the values of $U_{iso}$ and the temperature evolutions are similar, which suggests that local disorders are not largely increased by 50% substitution of Sn for the Pb site. In contrast, for $Ag_{1/3}Pb_{1/3}Bi_{1/3}Te$ ($\Delta S_{mix} = 1.10R$) and $Ag_{0.2}In_{0.2}Sn_{0.2}Pb_{0.2}Bi_{0.2}Te$ ($\Delta S_{mix} = 1.61R$), the values of $U_{iso}$ are clearly larger than that for PbTe. In addition, the extrapolation of the temperature dependence of $U_{iso}$ does not reach zero at 0 K for $Ag_{1/3}Pb_{1/3}Bi_{1/3}Te$) and $Ag_{0.2}In_{0.2}Sn_{0.2}Pb_{0.2}Bi_{0.2}Te$, which suggests unusual vibration and/or the presence of local disorder. These facts suggest that local disorders are created in MTe by M-site alloying with three or five element solutions. We find that the M-site $U_{iso}$ is larger than that of the Te site. The trend indicates that the displacements of M elements from average position are larger. In Figs. 3(e) and 3(f), $U_{iso}$s at $T = 100$ and 300 K are plotted as a function of $\Delta S_{mix}/R$ at the M site, respectively. At $T = 100$ K, the trend that $U_{iso}$ largely increases when M-site contains three or more elements in the solution is clear. For the data at $T = 300$ K, the observed trends are similar to those for $T = 100$ K, but Te-site $U_{iso}$ for $Ag_{0.2}In_{0.2}Sn_{0.2}Pb_{0.2}Bi_{0.2}Te$ is slightly smaller than that for $Ag_{1/3}Pb_{1/3}Bi_{1/3}Te$. The reason for the trend is not clear, but the possible origin would be glassy atomic vibrations, which will be discussed later, and/or local decompositions. Although SXRD patterns for $Ag_{0.2}In_{0.2}Sn_{0.2}Pb_{0.2}Bi_{0.2}Te$ are not broadened by heating (see Fig. S1(d) for data at $T =$ 400 K), we cannot exclude the possibility of local decomposition because the NaCl-type $Ag_{0.2}In_{0.2}Sn_{0.2}Pb_{0.2}Bi_{0.2}Te$ can be synthesized by a high-pressure annealing method only [23].

Next, we investigated atomic vibrations of NaCl-type MTe by MD simulation. Figure 4(a) shows the VDOS for MTe with M = Pb, $Sn_{0.5}Pb_{0.5}$, $Ag_{1/3}Sn_{1/3}Bi_{1/3}$ ($\Delta S_{mix}$ = 1.10$R$), $Ag_{0.25}Sn_{0.25}Pb_{0.25}Bi_{0.25}$ ($\Delta S_{mix}$ = 1.39$R$), $Ag_{0.2}In_{0.2}Sn_{0.2}Pb_{0.2}Bi_{0.2}$. For PbTe, many peaks corresponding to the vibration modes are observed, which indicates atomic vibrations in a normal crystalline compound. For M = $Sn_{0.5}Pb_{0.5}$, the peak structures can be seen, but the spectrum is broadened as compared to that for PbTe. In the spectrum for M = $Ag_{1/3}Sn_{1/3}Bi_{1/3}$, the VDOS peaks are not clearly seen, and the VDOS spectrum becomes flat in a wide frequency range. The broad spectrum suggests the suppression of characteristic vibration modes expected in a crystalline NaCl-type MTe compound. Here, we describe the flat VDOS spectrum as *glassy* vibrational characteristics, which means the presence of many vibrational modes in a disordered crystalline material with no atomic dispersions. The spectra for M = $Ag_{1/3}Sn_{1/3}Bi_{1/3}$ and $Ag_{0.25}Sn_{0.25}Pb_{0.25}Bi_{0.25}$ are quite similar except for the small differences at high frequencies, and the broadening of the spectrum is slightly enhanced in $Ag_{0.2}In_{0.2}Sn_{0.2}Pb_{0.2}Bi_{0.2}$. The broadening emerges with increasing band width; low and high frequency modes are excited. Similar phonon broadening was observed in alloy-based HEA [43]. However, the observed broadening in phonon dispersion in HEA was gradual when the number of solution element is increased from two to five. Therefore, the obvious modification of VDOS in MTe would be unique for compound-based HE materials. Figure 4(b) shows the temperature dependences of *dS* estimated as the difference in vibrational entropy at zero and the temperature. *dS* increases with increasing temperature, and larger *dS* is observed for MTe with a higher $\Delta S_{mix}$. The results indicate that vibrational entropy is increased with increasing $\Delta S_{mix}$ at the M site, and various local vibrational modes are induced, which results in glassy VDOS. We emphasize that the large affection of configurational entropy to atomic vibration occurs in the solution with $\Delta S_{mix} < 1.1R$ because the spectra for three data with $\Delta S_{mix} \geq 1.1R$ are almost the same. The trend is basically consistent with the results shown in Figs. 2 and 3, the robustness of superconductivity and the atomic displacements, and the glassy vibrations are created by the local disorder with a higher $\Delta S_{mix}$. In addition, we consider that the glassy VDOS will be observed in similar cubic CsCl-type structures as well, although simulation with CsCl-type model has not been done because all the systems become NaCl-type structure after structural relaxations.

Finally, we show the electronic structure of CsCl-type MTe, which is the high-$P$ phase where the robustness of superconductivity to pressure is observed. Figs. 5(a)–5(c) show the electronic band structure of CsCl-type MTe with M = Pb, $Ag_{0.33}Pb_{0.34}Bi_{0.33}$, $Ag_{0.2}In_{0.2}Sn_{0.2}Pb_{0.2}Bi_{0.2}$, respectively. For PbTe, sharp bands are observed. In contrast, blurry bands are observed for M = $Ag_{0.33}Pb_{0.34}Bi_{0.33}$, which is indicating the presence of local splitting of the electronic bands and band-shape modifications by the increase in $\Delta S_{mix}$. Noticeably, the obtained band structure for M = $Ag_{0.2}In_{0.2}Sn_{0.2}Pb_{0.2}Bi_{0.2}$ is quite similar to that of M = $Ag_{0.33}Pb_{0.34}Bi_{0.33}$, which suggests that an increase in $\Delta S_{mix}$ does not largely affect the electronic structure at an entropy range of $\Delta S_{mix} \geq 1.1R$, while a large modification of electronic structure is induced by an increase in local disorder at $\Delta S_{mix} < 1.1R$. In Fig. 5(d), the electronic DOS at Fermi energy ($E_F$) is plotted as a function of lattice constant, where the smaller lattice constant corresponds to higher applied pressure in CsCl-type MTe. The DOS basically decreases with pressure (with decreasing lattice constant). Therefore, conventional electron-phonon scenario can explain the decrease in $T_c$ in PbTe, but it cannot easily explain the behaviors in HE MTe with $\Delta S_{mix} \geq 1.1R$.

## 4. Discussion

We briefly discuss the origin of robustness of superconductivity under high pressures in CsCl-type HE MTe with the results reported here. On the basis of the BCS model (Bardeen-Cooper-Schrieffer model), a large electronic DOS, a high phonon frequency, and a larger electron-phonon coupling are preferable for a higher $T_c$ [44]. For PbTe, electronic DOS decreases with pressure, which explains the decrease in $T_c$ with pressure (Fig. 2(a)). As shown in Fig. 5(d), the electronic DOS for $Ag_{1/3}Pb_{1/3}Bi_{1/3}Te$ and $Ag_{0.2}In_{0.2}Sn_{0.2}Pb_{0.2}Bi_{0.2}Te$ decrease with pressure, but their $T_c$ does not decrease remarkably with pressure; $T_c$ for $Ag_{0.2}In_{0.2}Sn_{0.2}Pb_{0.2}Bi_{0.2}Te$ slightly increases with pressure, suggesting that the BCS scenario cannot be simply applicable in CsCl-type HE MTe. Figures 5(e) and 5(f) show the electronic DOS and the $T_c$ at two pressures, the lowest pressures for CsCl-type phases and $P = 30$ GPa, plotted as a function of $\Delta S_{mix}/R$. For PbTe and $Ag_{1/3}Pb_{1/3}Bi_{1/3}Te$, the difference in $T_c$ at the lowest pressure for the CsCl-type phase may be explained by the difference in DOS, but other trends cannot be understood with the electronic DOS only. Therefore, we consider that the $T_c$-$P$ trends cannot be simply understood by the electronic DOS viewpoints.

We propose that the possible scenario to explain the robustness of superconductivity in HE MTe is unique electron-phonon coupling with glassy phonon and electronic states. Furthermore, the glassy VDOS would be more important for the scenario because electronic DOS changes with pressure even in HE MTe as shown in Fig. 5(d). In addition, the degree of band blur is not remarkable near $E_F$ for $Ag_{0.33}Pb_{0.34}Bi_{0.33}Te$ and $Ag_{0.2}In_{0.2}Sn_{0.2}Pb_{0.2}Bi_{0.2}Te$ as shown in Fig. 5(b) and 5(c). It should be noted that the superconducting HE MTe phase is a metastable state, which makes it difficult to perform the detailed study using MD simulation. To obtain experimental evidences on the scenario proposed here, single crystals of HE MTe, particularly $Ag_{1/3}Pb_{1/3}Bi_{1/3}Te$ (or $Ag_{1/3}Sn_{1/3}Bi_{1/3}Te$) and $Ag_{0.2}In_{0.2}Sn_{0.2}Pb_{0.2}Bi_{0.2}Te$, with both NaCl-type and CsCl-type structures are desired. Furthermore, experimental investigations on the electronic structure (angle resolved photoemission spectroscopy) and phonons (inelastic neutron scattering) on those crystals are needed. Further studies on HE materials with stable state will be important to evidence our scenario of superconductivity in materials with glassy phonons and electronic structures and to apply the concept to developments of new HE functional materials.

In conclusion, we have investigated superconducting properties under high pressures for MTe with different M-site $\Delta S_{mix}$ ($0 \leq \Delta S_{mix} \leq 1.61R$) and confirmed that the robustness of superconductivity to pressure is induced by the increase in $\Delta S_{mix}$. From synchrotron XRD, we estimated the temperature and $\Delta S_{mix}$ dependences of $U_{iso}$ for those MTe and found that local structural disorder is induced when three or more elements are solved at the M site; in other words, local disorder is increased when $\Delta S_{mix}$ becomes greater than $1.1R$. The simulated VDOS spectra are broadened with increasing $\Delta S_{mix}$, and that for $1.1R \leq \Delta S_{mix} \leq 1.61R$ is quite similar. In the calculated band structures, we observed similar blurry bands for $Ag_{0.33}Pb_{0.34}Bi_{0.33}Te$ ($\Delta S_{mix} = 1.10R$) and $Ag_{0.2}In_{0.2}Sn_{0.2}Pb_{0.2}Bi_{0.2}Te$ ($1.61R$). Those results suggest that M-site alloying with three or more elements result in local structural disorders that create glassy VDOS and blurry electronic structure. Although experimental investigations with single crystals on the electronic and phonon characteristics in HE MTe with different $\Delta S_{mix}$ are needed to evidence the scenario, we here propose unique electron-phonon coupling with glassy phonon and electronic states to be the possible origin for the robustness of superconductivity to pressure in HE MTe. Our results shown here would be useful for designing HE materials including, exotic superconductors, high-$ZT$ thermoelectric materials, and other novel electronic and thermal functional materials.

**Acknowledgements**

Y. M. was partly supported by Grant-in-Aid for Scientific Research (KAKENHI) (No. 21H00151) and Tokyo Metropolitan Government Advanced Research (No. H31-1). R. K. was supported by JSPS KAKENHI (20H01874). K. T. was supported by JSPS KAKENHI (20H05619).

**Figures**

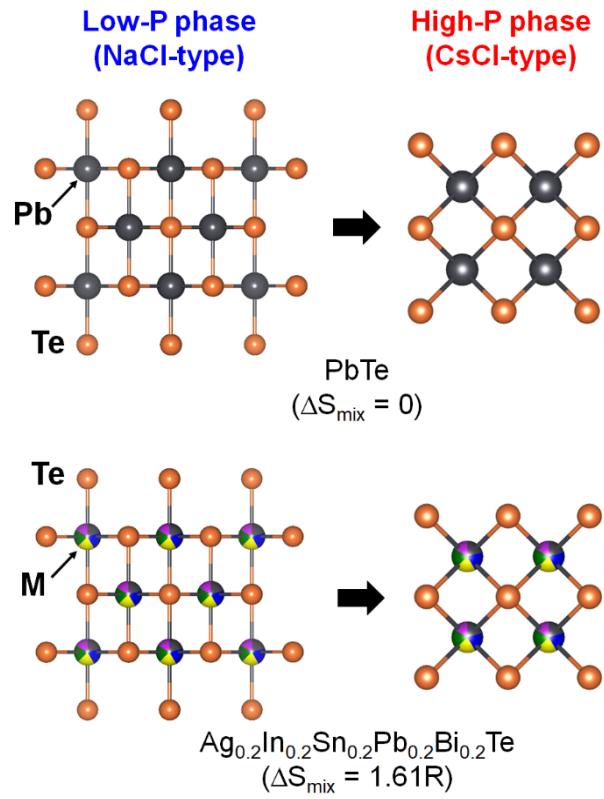

**Fig. 1. Schematic images of crystal structures.** Low-pressure (NaCl-type) and high-pressure (CsCl-type) crystal structures of PbTe and $Ag_{0.2}In_{0.2}Sn_{0.2}Pb_{0.2}Bi_{0.2}Te$ are displayed.

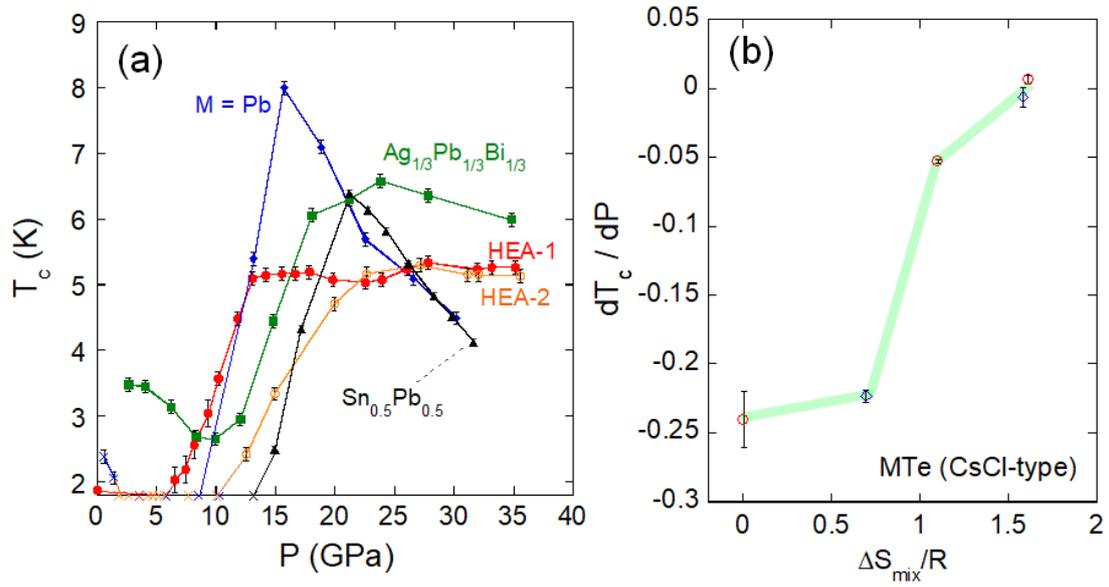

**Fig. 2. Robustness of superconductivity to pressure.** (a) Pressure dependences of $T_c$ for MTe. HEA-1 and HEA-2 denote samples with M = $Ag_{0.2}In_{0.2}Sn_{0.2}Pb_{0.2}Bi_{0.2}$ and $Ag_{0.175}In_{0.3}Sn_{0.175}Pb_{0.175}Bi_{0.175}$, respectively. Data for M = Pb, $Ag_{1/3}Pb_{1/3}Pb_{1/3}$, $Ag_{0.2}In_{0.2}Sn_{0.2}Pb_{0.2}Bi_{0.2}$ are reproduced from our previous paper [28]. The cross symbols indicate data where no superconducting transition was observed above 1.8 K. (b) $\Delta S_{mix}/R$ dependence of pressure differential of $T_c$ ($dT_c/dP$) in the CsCl-type structure. Light green lines are eye-guide.

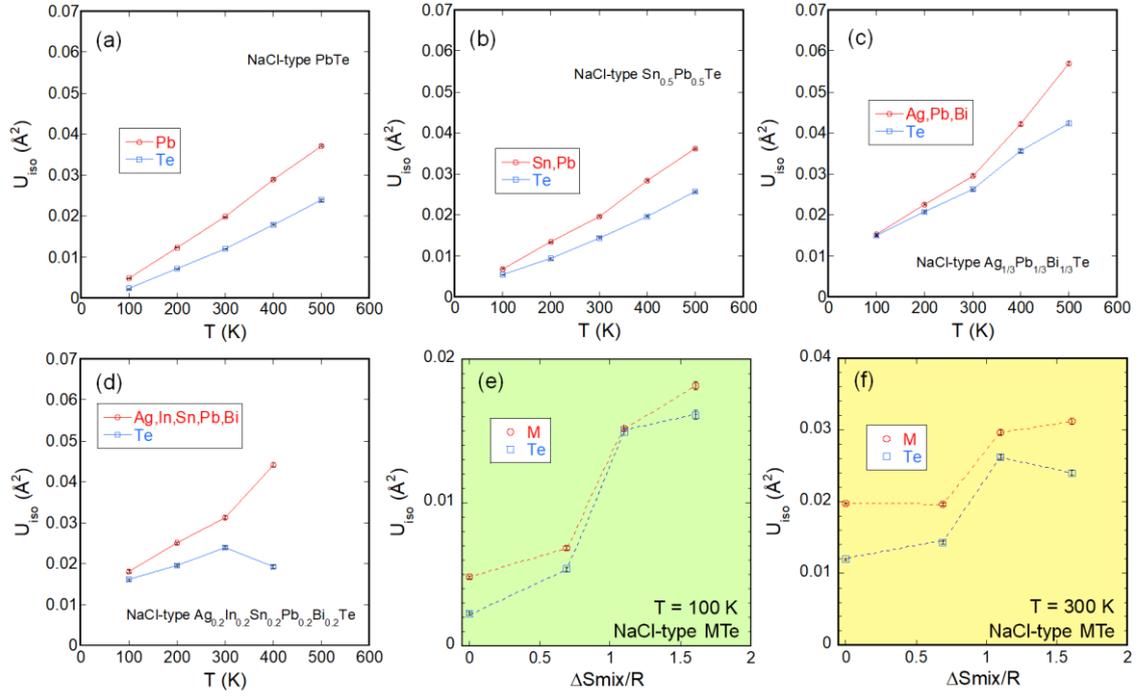

**Fig. 3. Atomic displacement parameters.** (a–d) Temperature dependences of $U_{iso}$ for M and Te sites for M = Pb, $Sn_{0.5}Pb_{0.5}$, $Ag_{1/3}Pb_{1/3}Pb_{1/3}$, $Ag_{0.2}In_{0.2}Sn_{0.2}Pb_{0.2}Bi_{0.2}$. (e,f) $\Delta S_{mix}/R$ dependences of $U_{iso}$ at $T$ = 100 and 300 K.

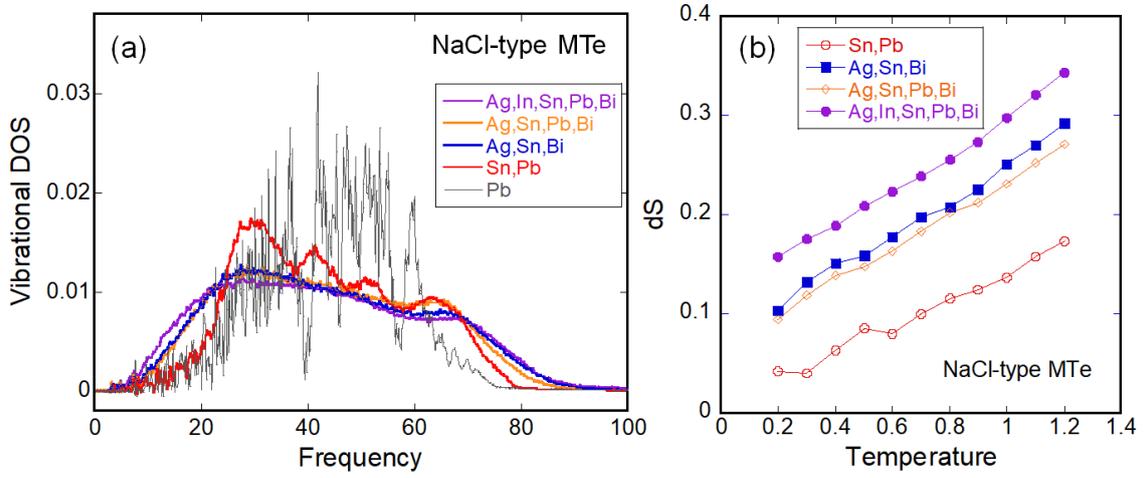

**Fig. 4. Simulated vibrational characteristics.** (a) Simulated VDOS spectra for MTe with M = Pb, $Sn_{0.5}Pb_{0.5}$, $Ag_{1/3}Sn_{1/3}Pb_{1/3}$, $Ag_{0.25}Sn_{0.25}Pb_{0.25}Bi_{0.25}$, $Ag_{0.2}In_{0.2}Sn_{0.2}Pb_{0.2}Bi_{0.2}$. (b) Temperature dependences of *dS* (entropy difference) for M = $Sn_{0.5}Pb_{0.5}$, $Ag_{1/3}Sn_{1/3}Pb_{1/3}$, $Ag_{0.25}Sn_{0.25}Pb_{0.25}Bi_{0.25}$, $Ag_{0.2}In_{0.2}Sn_{0.2}Pb_{0.2}Bi_{0.2}$. In the simulations, the parameters are normalized.

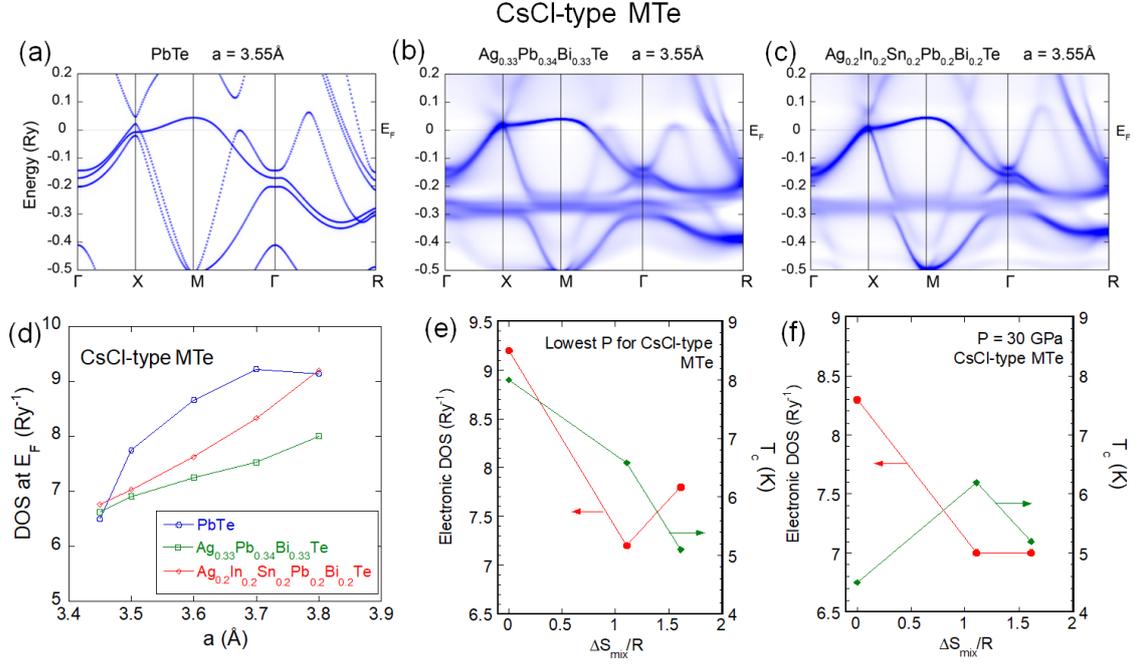

**Fig. 5. Calculated electronic structures.** (a–c) Electronic band structures for M = Pb, $Ag_{0.33}Pb_{0.34}Bi_{0.33}$, $Ag_{0.2}In_{0.2}Sn_{0.2}Pb_{0.2}Bi_{0.2}$. (d) Lattice constant dependences of electronic DOS. (e,f) $\Delta S_{mix}/R$ dependences of DOS and $T_c$ at pressures of (e) lowest pressures for the CsCl-phase and (f) $P = 30$ GPa.

# Supporting Information

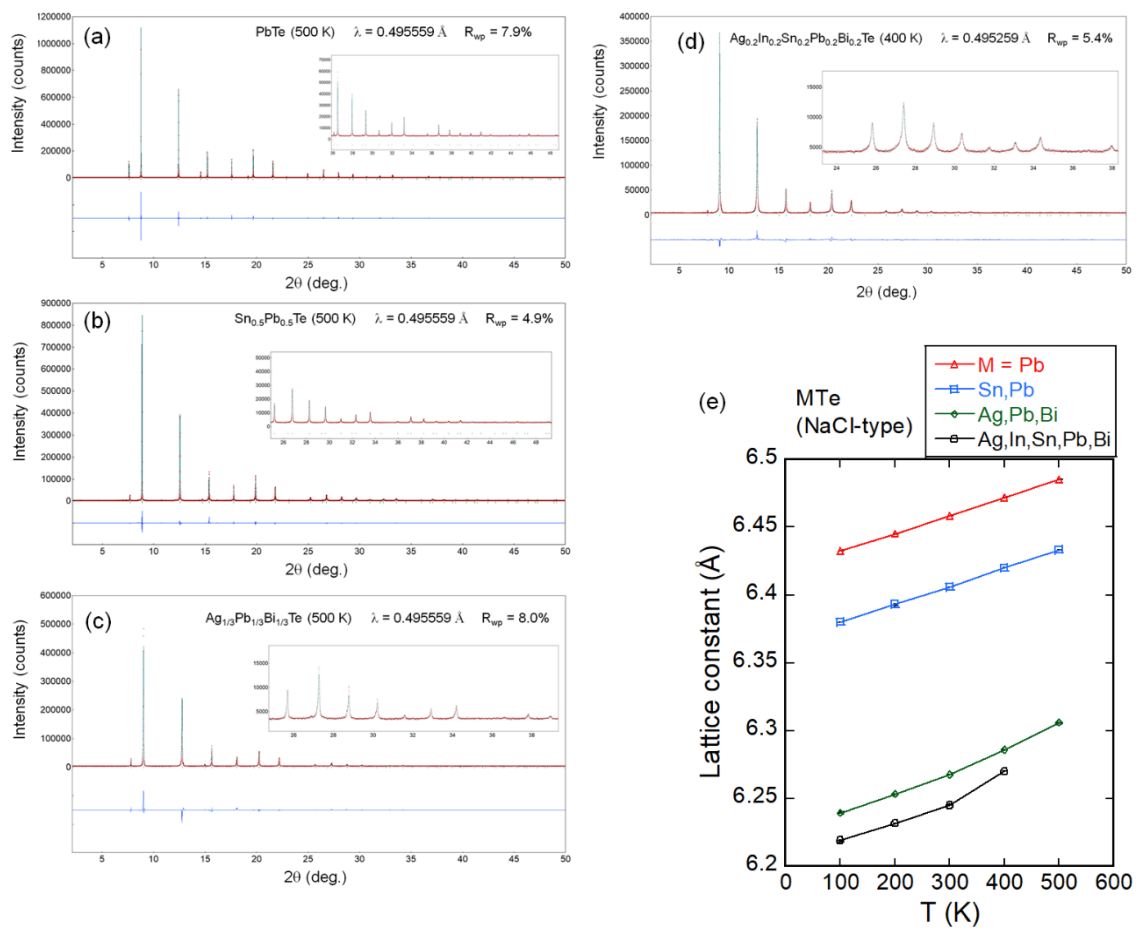

Fig. S1. (a–d) Synchrotron powder X-ray diffraction patterns and Rietveld refinement results for examined metal tellurides with NaCl-type structure. (e) Temperature dependences of lattice constant.

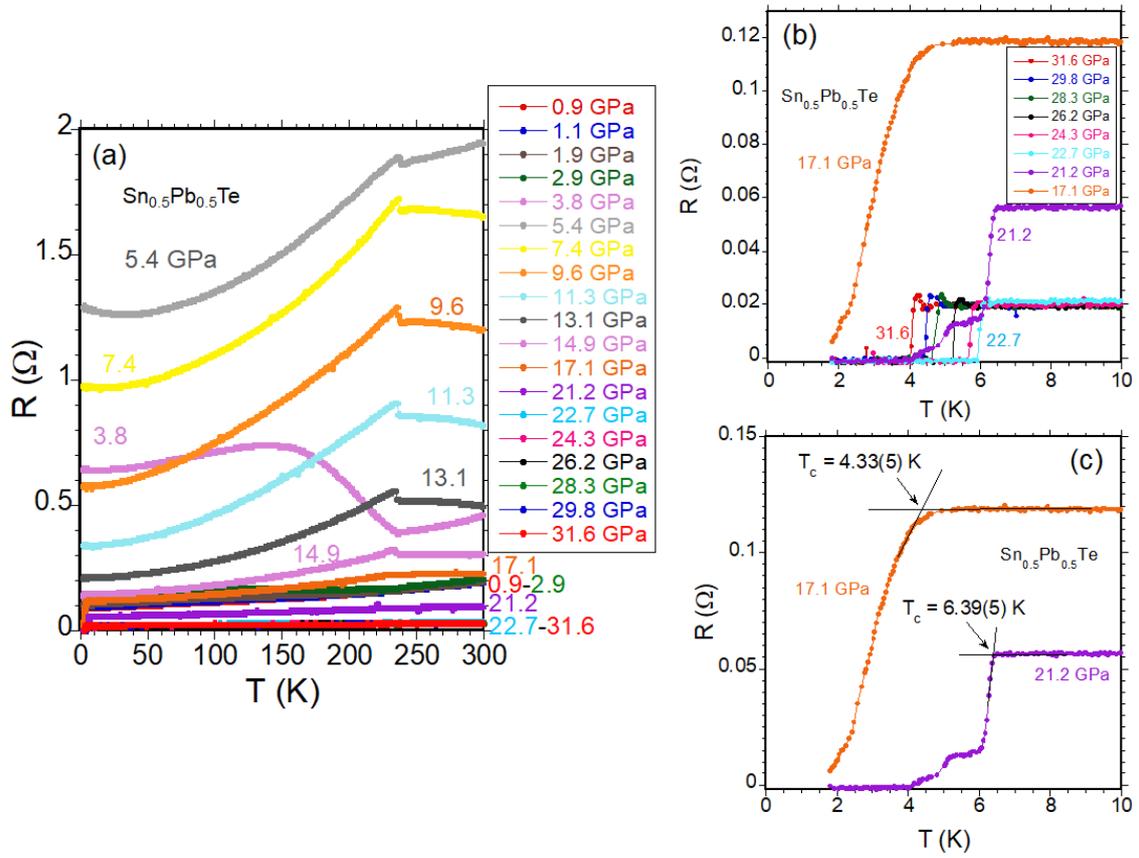

Fig. S2. (a) Temperature dependences of electrical resistance ($R$) for $Sn_{0.5}Pb_{0.5}Te$. (b) Low-$T$ resistance at high pressures above 17.1 GPa. (c) Examples of estimation of $T_c$ (onset of superconducting transition).

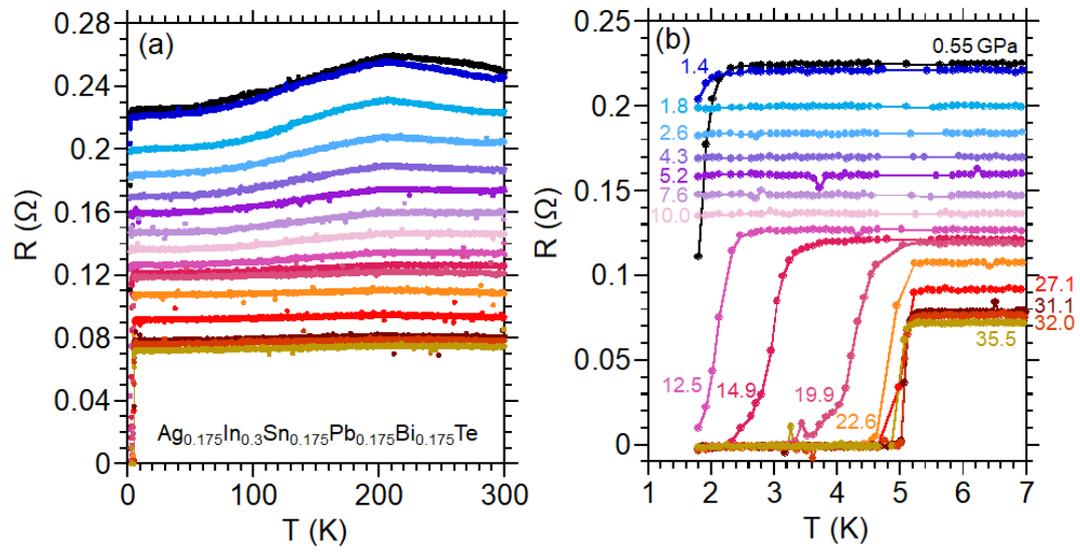

Fig. S3. (a) *R-T* for Ag$_{0.175}$In$_{0.3}$Sn$_{0.175}$Pb$_{0.175}$Bi$_{0.175}$Te. (b) *R-T* at low temperatures.

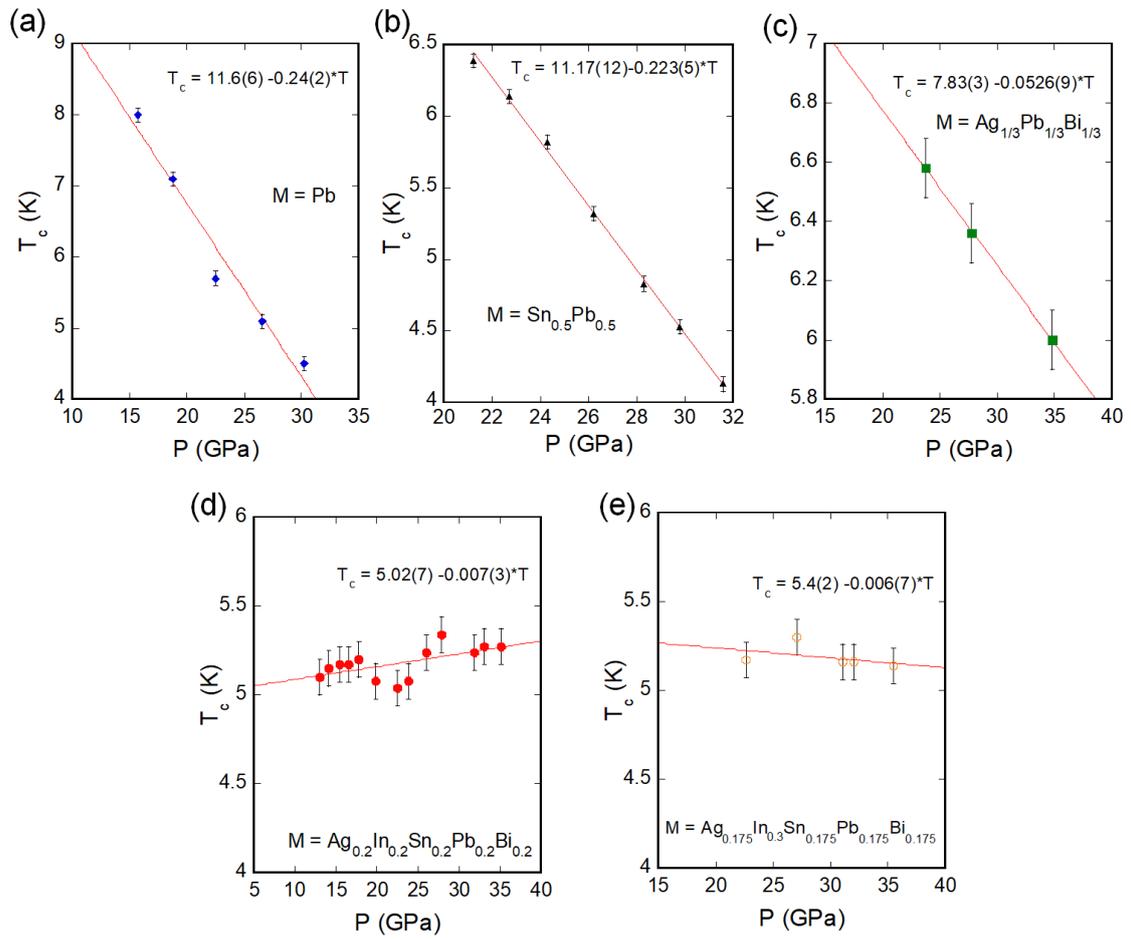

Fig. S4. Estimation of d$T_c$/d$P$ from the data of pressure dependence of $T_c$. Here linear fitting was performed and the result formula is displayed. $T_c$ data in (a,c,d) were taken from our previous publication (Md. R. Kasem et al., *Sci. Rep.* 2022, **12**, 7789.).